# An Extensive Photometric Investigation of theW UMa System DK Cyg

M. M. Elkhateeb[1,2] and M. I. Nouh[1,2], E. Elkholy[1,2] and B. Korany[1,3]
abdo_nouh@hotmail.com

[1]Astronomy Department, National Research Institute of Astronomy and Geophysics, 11421, Helwan, Cairo, Egypt.

[2]Physics department, College of Science, Northern Border University, 1321 Arar, Saudi Arabia

[3]Physics Department, Faculty of Applied Science, Umm Al-Qura University, Makkah 715, Saudi Arabia

**Abstract:** DK Cyg (P = 0.4707) is a W Ursae Majoris type eclipsing binary system that undergoes complete eclipses. All the published photoelectric data have been collected and utilized to re-examine and update the period behavior of the system. A significant period increase with rate of $12.590 \times 10^{-11}$ day/cycle was calculated. New period and ephemeris has been calculated for the system. A long term photometric solution study was performed and a light curve elements were calculated. We investigated the evolutionary status of the system using theoretical evolutionary models.

## 1. Introduction

The eclipsing binary DK Cyg (BD +330 4304, 10.37-10.93 $m_V$ and spectral type A7V) is a well known W UMa type binary with a period of about 0.4707 days. It was discovered as variable star earlier by Guthnick&Prager (1927), so their epoch of intensive observations is very long. The earliest photographic light curve shows that the system is of the W Ursa Majoris type. Visual light curves were published by Piotrowski (1936) and Tsesevich (1954) from Klepikova's observations.

First photoelectric observations for the system were carried out by Hinderer (1960), while Binnendijk (1964) observed the system photoelectrically in B and V bands and derived least squares orbital solution using his observed light curves. The observations showed scatter which may indicates non-periodic fluctuations, especially in the secondary minimum. The system DK Cyg was classified in the general Catalogue of Variable Stars as A7V (Samus, 2014), while Binnendijk (1964) adopted it as A2. Mochnacki&Doughty (1972) showed that the color index of the system judged its spectral type and found that the spectral type of the system is more likely to be about F0 to F2. Because of the system DK Cyg is a summer object in the Northern hemisphere with 11.5-hour period and short durations of night, it is bound to remain ill-observed (Awadalla, 1994). Only three complete light curves by Binnendijk (1964), Paparo et al. (1985), and Awadalla (1994) were published. Photoelectric observations and new times of minima have been carried by many authors: Borkovits et al. (2004), Sarounova and Wolf (2005), Drozdz and Ogloza (2005), Hubscher et al. (2006), Dogru et al. (2007), Hubscher et al. (2008), Hubscher et al. (2010), Erkan et al. (2010), Diethelm (2010), and Dogru et al. (2011) ,Simmons, (2011) ,Diethelm (2012) ,and Diethelm (2013).

In the present paper we are going to perform comprehensive photometric study of the WMa system DK Cyg.

## 2. Period change

Although the period variation of contact binary systems of the W UMa-type is a controversial issue of binary star astrophysics, the cause of the variations (long- as well as short-term) is still a mystery for a discussion of possible physical mechanisms (Kiss et al., 1999). Magnetic activity cycle is one of the main mechanisms that caused a period variation together with the mass exchange between the components of each system. Observations by Binnendijk (1964) showed a change of the secondary minimum depth and a new linear light element was derived. Period study by Paparo et al. (1985) showed that the orbital period of the system DK Cyg increases and the first parabolic light elements were calculated, which confirmed the light curve variability. Kiss et al. (1999) updated the linear ephemeris of DK Cyg, while Awadalla (1994) re-calculated a new quadratic element for the system and confirmed the light curve variability suggested by Paparo et al. (1985). Wolf et al. (2000) used a set of 101 published times of minimum covering the interval between 1926 – 2000 in order to update the quadratic element calculated by Awadalla (1994). They showed that the period increases by the rate $11.5 \times 10^{-11}$ day/cycle. Borkovits et al. (2005) follow the period behavior of the system using set of published minima from HJD 2424760 TO HJD 2453302. A continuous period increasing was adopted in agreement with previous results with an updated quadratic element.

In this paper we studied the orbital period behavior of the system DK Cyg using the (O-C) diagram based on more complete data set collected from the literatures and databases of BAV, AAVSO, and BBSAG observers. Part of our collected data set was given by Kreiner et al. (2001); and unpublished Hipparcos observations and main part were downloaded from website (http:J//astro.sci.muni.cz/variables/ocgate/); (see Table 1). A total of 195 minima times were incorporated in our analysis covering about 86 year (66689 orbital revolutions) from 1927 to 2013. It's clear that our set of data added about 94 of new minima and increases the interval limit of the orbital period study about 13 year over than the data of Wolf et al. (2000), which give an accurate indication about the period behavior of the system. The different type of collected minima ( i.e. photographic, visual, photoelectric and CCD) were weighted according to their type. The residual (O-C)'s were computed using Binnendijk (1964) ephemeris (Eq.1), and represented in Fig. 1 versus the integer cycle E, no distinction has been made between primary and secondary minima.

$$\text{Min I} = 2437999.5838 + 0.47069055 * E \qquad (1)$$

It can be seen from the figure that the behavior of the orbital period of the system DK Cyg shows a parabolic distribution which generally interpreted by the transfer of mass from one component to the other of binary. Reasonable linear least-squares fit of the data available improved the light elements given in Eq. (1) to:

$$\text{Min I} = 2437999.5961 + 0^d.47069206 * E \qquad (2)$$
$$\pm 0.0243 \quad \pm 0.0192$$

The linear element yields a new period of $P = 0^d.47069206$ day which is longer by 0.13 second with respect to the value given by Binnendijk (1964). Quadratic least-squares fit gives:

$$\text{Min I} = 2437999.5803 + 0^d.47069064 * E + 6.284*10^{-11} * E^2 \qquad (3)$$
$$\pm 0.0192 \quad \pm 0.0237 \quad \pm 2.5654 \times 10^{-12}$$

The rate of period increasing resulting from the quadratic elements (Eq.3) is $dP/dE = 12.568 \times 10^{-11}$ day/cycle or $9.746 \times 10^{-8}$ day/year or 0.84 second/century. As the (O-C) curve shows a parabolic distribution, we can refer the period variation to the mass transfer between the components of the system. More future systematic and continuous photometric observations are needed to follow a continuous change in the orbital period of the system DK Cyg which may show a periodic behavior. The fourth column of Table 1 represents the quadratic residuals (O-C)q calculated using the new element of Eq. 2 and represented in Fig. 2. All published linear and quadratic elements together with that result from our calculations are listed in Table 2. It is noted from the table that the quadratic term resulted from our calculations have slightly higher value than that calculated by Awadalla (1994), Wolf et al. (2000), and Borkovits et al. (2005), which can be interpreted by the increases the set of minima in our study than that they used (nearly double) and also we covered an interval larger than that they used.

Table 1: Times of minimum light for DK Cyg.

| HJD | Method | E | (O-C) | (O-C)q | Ref | HJD | Method | E | (O-C) | (O-C)q | Ref |
|---|---|---|---|---|---|---|---|---|---|---|---|
| 2434179.4690 | vis | -8116 | 0.00970 | 0.00971 | 1 | 2451749.4450 | pe | 29212 | 0.04885 | -0.00370 | 3 |
| 2447758.4352 | pe | 20733 | 0.02423 | -0.00099 | 2 | 2451777.6740 | ccd | 29272 | 0.03642 | -0.01636 | 5 |
| 2447790.4437 | pe | 20801 | 0.02577 | 0.00037 | 2 | 2452163.6590 | ccd | 30092 | 0.05517 | -0.00074 | 5 |
| 2447963.6620 | pe | 21169 | 0.02995 | 0.00354 | 3 | 2452245.5592 | ccd | 30266 | 0.05521 | -0.00137 | 5 |
| 2447963.8960 | pe | 21169.5 | 0.02860 | 0.00220 | 3 | 2452253.5613 | ccd | 30283 | 0.05557 | -0.00108 | 5 |
| 2448265.1380 | pe | 21809.5 | 0.02865 | 0.00046 | 4 | 2452441.8384 | ccd | 30683 | 0.05645 | -0.00176 | 5 |
| 2448265.1382 | pe | 21809.5 | 0.02885 | 0.00066 | 4 | 2452512.4415 | pe | 30833 | 0.05597 | -0.00284 | 7 |
| 2448272.1987 | pe | 21824.5 | 0.02899 | 0.00076 | 4 | 2452525.6231 | ccd | 30861 | 0.05824 | -0.00069 | 5 |
| 2448297.6160 | pe | 21878.5 | 0.02900 | 0.00062 | 3 | 2452526.5644 | ccd | 30863 | 0.05816 | -0.00078 | 5 |
| 2448302.7930 | pe | 21889.5 | 0.02841 | -0.00001 | 3 | 2452811.8062 | ccd | 31469 | 0.06148 | 0.00013 | 5 |
| 2448308.2078 | pe | 21901 | 0.03027 | 0.00182 | 4 | 2453223.4286 | ccd | 32343.5 | 0.06500 | -0.00006 | 8 |
| 2448308.2079 | pe | 21901 | 0.03037 | 0.00192 | 4 | 2453228.3681 | ccd | 32354 | 0.06225 | -0.00274 | 8 |
| 2448336.4491 | pe | 21961 | 0.03013 | 0.00152 | 4 | 2453246.4950 | ccd | 32392.5 | 0.06756 | 0.00242 | 8 |
| 2449988.5840 | ccd | 25471 | 0.04120 | 0.00182 | 5 | 2453247.4346 | ccd | 32394.5 | 0.06578 | 0.00063 | 8 |
| 2450003.6456 | ccd | 25503 | 0.04070 | 0.00122 | 5 | 2453285.3260 | ccd | 32475 | 0.06659 | 0.00110 | 8 |
| 2450313.8240 | ccd | 26162 | 0.03403 | -0.00765 | 5 | 2453286.2657 | ccd | 32477 | 0.06491 | -0.00059 | 8 |
| 2450341.6130 | ccd | 26221 | 0.05229 | 0.01041 | 5 | 2453302.2672 | ccd | 32511 | 0.06293 | -0.00271 | 8 |
| 2450397.6060 | ccd | 26340 | 0.03311 | -0.00917 | 5 | 2454799.5505 | ccd | 35692 | 0.07959 | 0.00004 | 9 |
| 2450692.7400 | ccd | 26967 | 0.04414 | -0.00030 | 5 | 2455043.8381 | ccd | 36211 | 0.07882 | -0.00310 | 10 |
| 2451000.0990 | pe | 27620 | 0.04221 | -0.00452 | 5 | 2455062.6680 | ccd | 36251 | 0.08107 | -0.00105 | 11 |
| 2451095.6600 | ccd | 27823 | 0.05303 | 0.00557 | 5 | 2455088.5544 | ccd | 36306 | 0.07953 | -0.00285 | 10 |
| 2451160.5980 | ccd | 27961 | 0.03573 | -0.01222 | 5 | 2455810.6029 | ccd | 37840 | 0.08870 | -0.00096 | 12 |
| 2451379.4820 | pe | 28426 | 0.04863 | -0.00101 | 3 | | | | | | |

**References:** 1- Szafraniec (1953); 2- Hubscher et al. (1990); 3- Wolf et al. (2000); 4- Hipparcos observations (unpublished); 5- Baldwin&Samolyk (2003); 6- Kiss et al. (1999); 7- Borkovits et al. (2002); 8- Borkovits et al. (2005); 9- Gerner (2008); 10- Menzies (2009); 11- Samolyk (2009); 12- Simmons (2011).

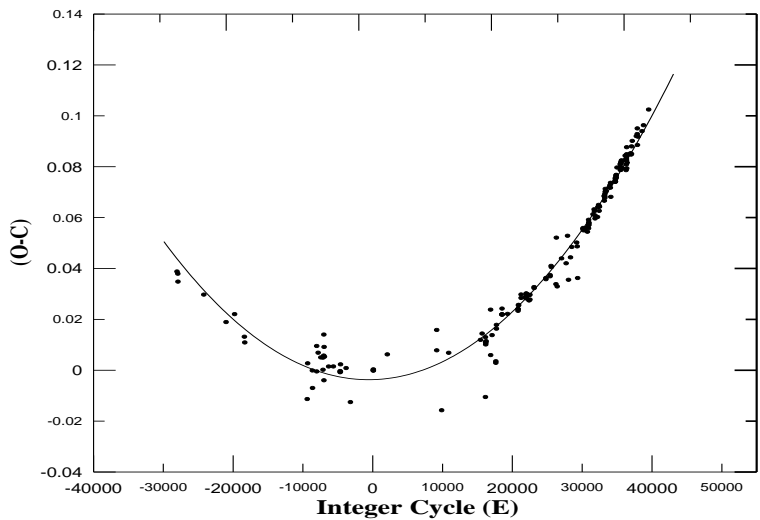

Figure 1: Period behavior of DK Cyg.

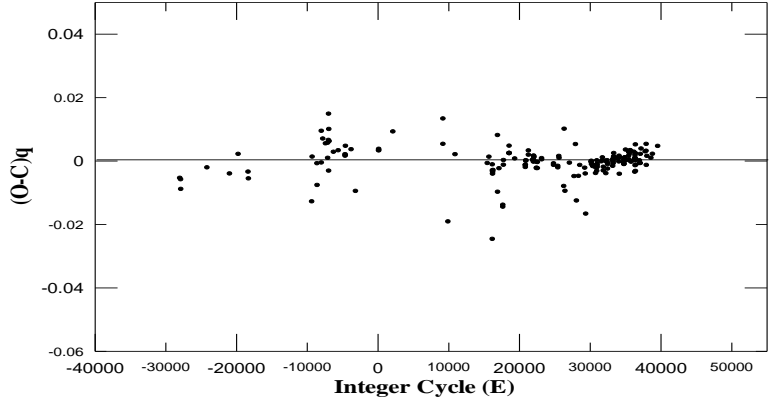

Figure 2: Calculated residuals from the polynomial ephemeris.

Table 2: The light elements of DK Cyg.

| JD | Period | Quadratic term | Ref. |
|---|---|---|---|
| 2437999.5838 | 0.470690550 | | Binnendijk (1964) |
| 2437999.5828 | 0.470690660 | $5.390 \times 10^{-10}$ | Paparo et al. (1985). |
| 2437999.5825 | 0.470690730 | $5.760 \times 10^{-11}$ | Awadalla (1994) |
| 2451000.0999 | 0.470692900 | | Kiss et al. (1999) |
| 2437999.5825 | 0.470690640 | $5.750 \times 10^{-11}$ | Wolf et al. (2000) |
| 2451000.1031 | 0.470693909 | $5.862 \times 10^{-11}$ | Borkovits et al. (2005) |
| 2437999.5961 | 0.470692060 | | Present work |
| 2437999.5803 | 0.470690640 | $6.284 \times 10^{-11}$ | Present work |

## 3. Light Curve Modeling

Light curve modeling for the system DK Cyg by Mochnacki&Doughty (1972) using Binnendijk (1964) observations in V band showed un-matching between the theoretical curve and the observations. The theoretical curve is deeper than the observations at primary minimum, which may give inaccurate light curve parameters. The photometric mass ratio calculated from their accepted model was $q_{ph}=0.33\pm0.02$, while the spectroscopic value estimated using radial velocity study by Rucinski and Lu (1999) is $q_{sp}=0.32\pm0.04$. In the present work we used the complete published light curves by Binnendijk (1964), Paparo et al. (1985), and Awadalla (1994) in V-band through a long term photometric solution study in order to estimate the physical parameters of the system, to follow its evolutionary status. Light curves published by Baran et al. (2004) didn't included in our study because the individual observations of the observed light curves are not avilable for us.

Photometric analysis for the studied light curves of the system DK Cyg were carried out using Mode 3 (overcontact) of WDint56a Package (Nelson 2009) which based on the 2009 version of Willson and Devinney (W-D) code. The observed light curves were analyzed using all individual observations instead of the normal points which don't reveal a real light variation of the system. Gravity darkening and bolometric albedo exponents appropriate were assumed for the convective envelopes. We adopted $g_1 = g_2 = 0.32$ (Lucy 1967) and $A_1 = A_2 = 0.5$ (Rucinski, 1969). Bolometric limb darkening values are adopted using the table of Van Hamme (1993). Temperature of the primary star was adopted according to Baran et al. (2004) model with some amendment according to the studied light.

The adjustable parameters are the mean temperature of the secondary component $T_2$, orbital inclination $i$, and the potential of the two components $\Omega = \Omega_1 = \Omega_2$, while the spectroscopic mass ratio ($q_{sp}= 0.32 \pm 0.04$) by Rucinski and Lu (1999), was fixed for all calculated models together with the primary star's temperature ($T_1$). Photometric solutions for the studied light curves using unspotted models (not shown here) do not fit the observed ones well at all. Therefore, a spotted model was adopted and a photometric fitting was reached after several runs, which showing a well matching between the theoretical light cures and the observations. Table 3 lists the calculated parameters for the three light curves, while Figure 3 represented the theoretical light curves according to the accepted solution together with the reflected points in V-band. The $\sum (O-C)^2$ in Table 3 are indicative of comparisons in future studies, since the number of observations and the accuracy is not the same in the three light curves. The accepted solutions revealed a spotted model with migration in the spots longitude and decrease in spot radius, which mean that our spotted model give a reasonable uniform description of the system DK Cyg. The accepted model using Awadalla (1994) light curve, including additional hot spot (spot *B*) in the primary component which is exposed during secondary eclipse causes raising in Min II, and interprets the sudden increase in the light level at secondary minimum as noted by Awadalla (1994). Absolute physical parameters for each component of the system DK Cyg were calculated based on the results of the radial velocity data of Baran et al. (2004) and our new photometric solution for each light curve. The calculated parameters are listed in Table 3. The results show that the primary component is more massive and hotter than the secondary component. A three dimensional geometrical structure for the system DK Cyg is displayed in Figure 4 using the software Package Binary Maker 3.03 (Bradstreet and Steelman, 2002) based on the calculated parameters resulted from our models.

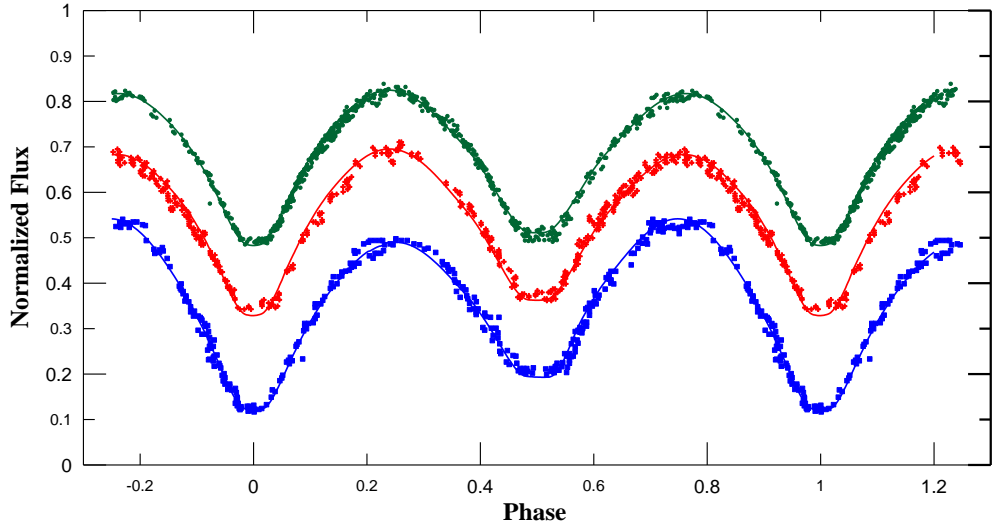

Figure 3: Observed and synthetic light curves of Binndijik (1964), Paparo et al. (1985), and Awadalla (1994) for the system DK Cyg.

Table 3: Photometric solution for DK Cyg.

| Parameter | Binnendijk (1964) | Paparo et al (1985) | Awadalla (1994) |
|---|---|---|---|
| $A$ | 5500 | 5500 | 5500 |
| $i(^0)$ | 80.55±0.12 | 80.88±0.29 | 82.27±0.37 |
| $g_1 = g_2$ | 0.32 | 0.32 | 0.32 |
| $A_1 = A_2$ | 0.5 | 0.5 | 0.5 |
| $q\ (M_2 / M_1)$ | 0.322* | 0.322* | 0.322* |
| $\Omega_1 = \Omega_2$ | 2.4317±0.003 | 2.4333±0.005 | 2.3941±0.004 |
| $\Omega_{in}$ | 2.5143 | 2.5143 | 2.5143 |
| $\Omega_{out}$ | 2.3225 | 2.3144 | 2.3144 |
| $T_1\ (^oK)$ | 7400* | 7400* | 7400* |
| $T_2\ (^oK)$ | 6632 ±5 | 6613±10 | 6629±10 |
| $L_1/(L_1+L_2)$ | 0.8057 | 0.8056 | 0.7996 |
| $L_2/(L_1+L_2)$ | 0.1944 | 0.1944 | 0.2004 |
| $r_1$ pole | 0.4673±0.0007 | 0.4670±0.0014 | 0.4755±0.0011 |
| $r_1$ side | 0.5064±0.0010 | 0.5059±0.0019 | 0.5179±0.0016 |
| $r_1$ back | 0.5384±0.0013 | 0.5378±0.0025 | 0.5538±0.0022 |
| $r_2$ pole | 0.2848±0.0008 | 0.2844±0.0015 | 0.2939±0.0013 |
| $r_2$ side | 0.2994±0.0010 | 0.2989±0.0018 | 0.3108±0.0016 |
| $r_2$ back | 0.3495±0.0019 | 0.3485±0.0037 | 0.3742±0.0039 |
| Spot $A$ of star **1** | | | |
| Co-latitude | 130* | 130* | 130* |
| Longitude | 180* | 180* | 180* |
| Spot radius | 33.78±0.230 | 27.97±0.316* | 22.35±0.40 |
| Temp. factor | 0.821±0.003 | 0.708±0.014* | 0.71±0.02 |
| Spot *B* of star **1** | | | |
| Co-latitude | | | 90* |
| Longitude | | | 198* |
| Spot radius | | | 14.58±0.56 |
| Temp. factor | | | 1.088±0.01 |
| Spot *A* of star **2** | | | |
| Co-latitude | 120* | 120* | 120* |
| Longitude | 290* | 250* | 290* |
| Spot radius | 28.44±0.901 | 29.42±1.34 | 29.41±0.45 |
| Temp. factor | 0.917±0.014 | 0.882±0.02 | 1.33±0.01 |
| $\sum (O-C)^2$ | 0.03033 | 0.05063 | 0.03043 |

*Not adjusted

Table 4: Absolute physical parameters for DK Cyg.

| Parameter | Binnendijk (1964) | Paparo et al. (1985) | Awadalla (1994) |
|---|---|---|---|
| $M_{1\odot}$ | 1.7212±0.070 | 1.7199±0.070 | 1.7117±0.070 |
| $M_{2\odot}$ | 0.5542±0.023 | 0.5538±0.023 | 0.5512±0.023 |
| $R_{1\odot}$ | 1.6880±0.069 | 1.6860±0.069 | 1.7240±0.070 |
| $R_{2\odot}$ | 1.0420±0.043 | 1.0400±0.043 | 1.0900±0.045 |
| $T_{1\odot}$ | 1.2810±0.052 | 1.2810±0.052 | 1.2810±0.052 |
| $T_{2\odot}$ | 1.1480±0.047 | 1.1450±0.006 | 1.1470±0.006 |
| $M_{1\_bol}$ | 2.5400±0.104 | 2.5430±0.104 | 2.4940±0.102 |
| $M_{2\_bol}$ | 4.0640±0.166 | 4.0800±0.167 | 3.9680±0.162 |
| $L_{1\odot}$ | 7.6550±0.313 | 7.6370±0.312 | 7.985±0.326 |
| $L_{2\odot}$ | 1.8820±0.077 | 1.8530±0.076 | 2.056±0.084 |

Note: subscript 1 and 2 means primary and secondary component respectively.

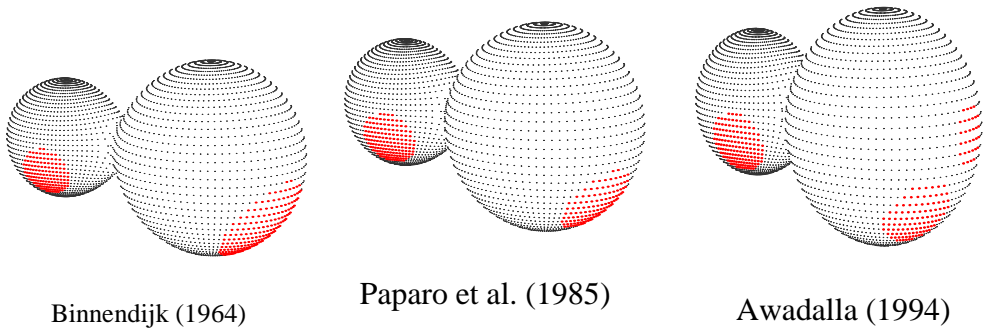

Binnendijk (1964)    Paparo et al. (1985)    Awadalla (1994)

Figure 4: Three-dimensional models of the components of DK Cyg.

## 4. Discussion and Conclusion

Studying of the period behavior of the system DK Cyg based all available published times of minima, covering 86 yr of observations including 195 times of light minima shows a continuous period increase with the rate $dP/dE = 12.590 \times 10^{-11}$ day/cycle or $9.763 \times 10^{-8}$ day/year or 0.84 second/century. A new linear and quadratic elements were calculated based on all available published data and yields a new period of $P = 0.47069203$ day. A long term photometric study was performed using published observations by Binnendijk (1964), Paparo et al. (1985), and Awadalla (1994). Absolute physical parameters for the system were calculated based on the accepted photometric solutions. More systematic and continuous photometric observations for the system DK Cyg are needed to confirm a continuous change in the period and follow its light curve variation.

One of the difficulties for W UMa binaries is to use stellar models of single stars to investigate the evolutionary status of these systems. However, using these theoretical models may give approximate view about the evolutionary status of the system.

We used the physical parameters listed in Table 4 to investigate the current evolutionary status of DK Cyg. In Figures 5 and 6, we plotted the components of DK Cygon the mass–luminosity (M-L) and mass–radius (M-R) relations along with the evolutionary tracks computed by Girardi et al. (2000) for both zero age main sequence stars (ZAMS) and terminal age main sequence stars (TAMS) with metalicity $z = 0.019$. As it is clear from the figures, the primary component of the system is located nearly on the ZAMS for both the M-L and M-R relations. The secondary component is above the TAMS track for M-L and the M-R relations. For the sake of comparison, we plotted sample of W-type contact binaries listed in Table 5. The components of DK Cyg have the same behavior of the selected W-type systems.

The mass-effective temperature relation (M–$T_{eff}$) relation for intermediate and low mass stars (Malkov, 2007) is displayed in Figure 7. The location of our mass and radius on the diagram revealed a good fit for the primary and poor fit for the secondary components. This gave the same behavior of the system on the mass-luminosity and mass-radius relations.

**Table 5: Physical parameters of the eight W-type contact binaries**

| Star Name | Parameter | | | | | | | | Ref. |
|---|---|---|---|---|---|---|---|---|---|
| | $M_1(M_\odot)$ | $M_2(M_\odot)$ | $R_1(R_\odot)$ | $R_2(R_\odot)$ | $L_1(L_\odot)$ | $L_2(L_\odot)$ | $T_1(T_\odot)$ | $T_2(T_\odot)$ | |
| TY Boo | 1.03 | 0.48 | 1.02 | 0.72 | 0.5 | 0.89 | 0.99 | 0.95 | 1 |
| AW UMa | 1.6 | 0.121 | 1.786 | 0.739 | 7.47 | 0.804 | 1.242 | 1.218 | 2 |
| AD Cnc | 0.93 | 0.58 | 0.89 | 0.72 | 0.39 | 0.33 | 0.804 | 0.860 | 3 |
| TX Cnc | 1.37 | 0.82 | 1.23 | 0.96 | 2.2 | 1.38 | 1.056 | 1.066 | 3 |
| RZ Com | 1.03 | 0.45 | 1.06 | 0.71 | 0.93 | 0.43 | 0.916 | 0.927 | 3 |
| LS Del | 1.06 | 0.60 | 1.09 | 0.83 | 1.12 | 0.69 | 0.950 | 0.963 | 3 |
| BB Peg | 1.16 | 0.47 | 1.21 | 0.78 | 1.58 | 0.81 | 0.980 | 1.033 | 3 |
| AA UMa | 1.26 | 0.69 | 1.40 | 1.10 | 2.17 | 1.43 | 0.988 | 1.005 | 3 |

**Reference: 1- Elkhateeb et al. (2014), 2- Elkhateeb and Nouh (2014), 3- Maceroni and van't Veer (1996).**

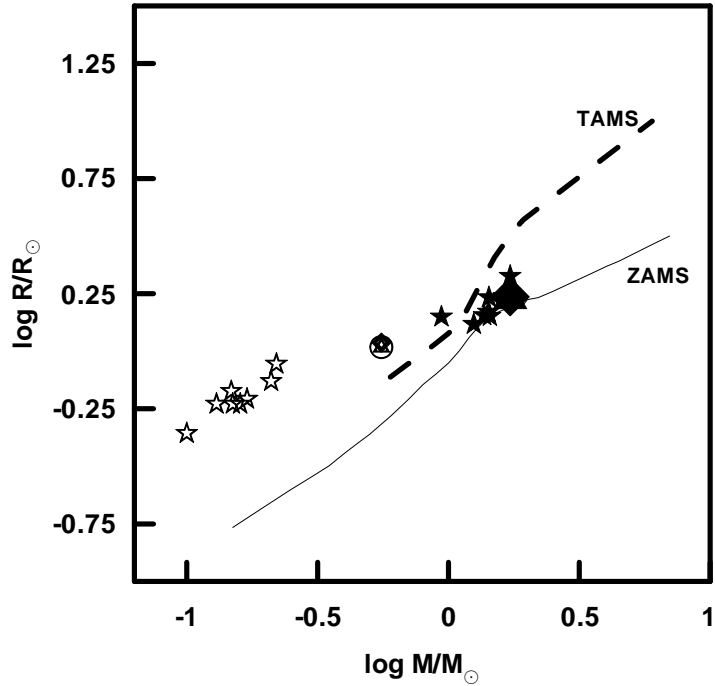

Figures 5: The position of the components of DK Cyg on the mass–radius diagram. The filled circle denotes the primary and the open circle represents the secondary. The other symbols denote the sample of the selected W-type systems listed in Table 5.

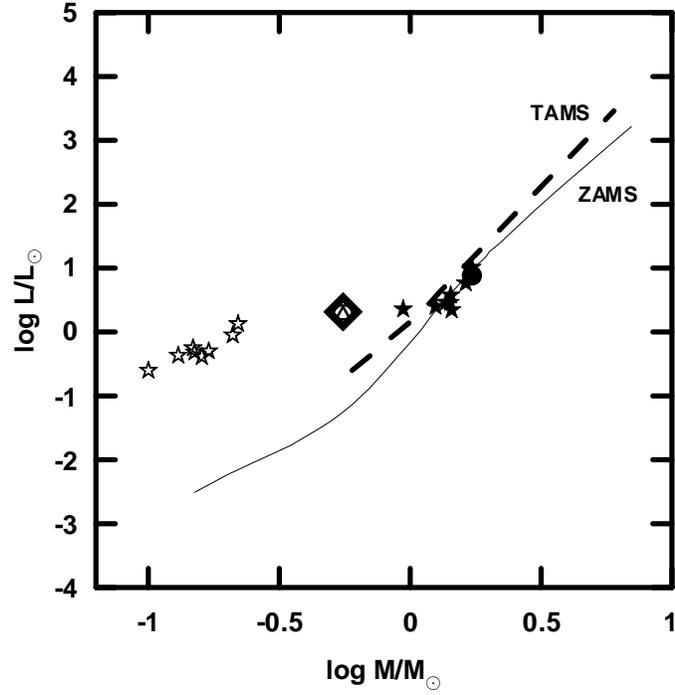
Figures 6: The position of the components of DK Cyg on the mass–luminosity diagram. The filled circle denotes the primary and the open circle represents the secondary. The other symbols denote the sample of the selected W-type systems listed in Table 5.

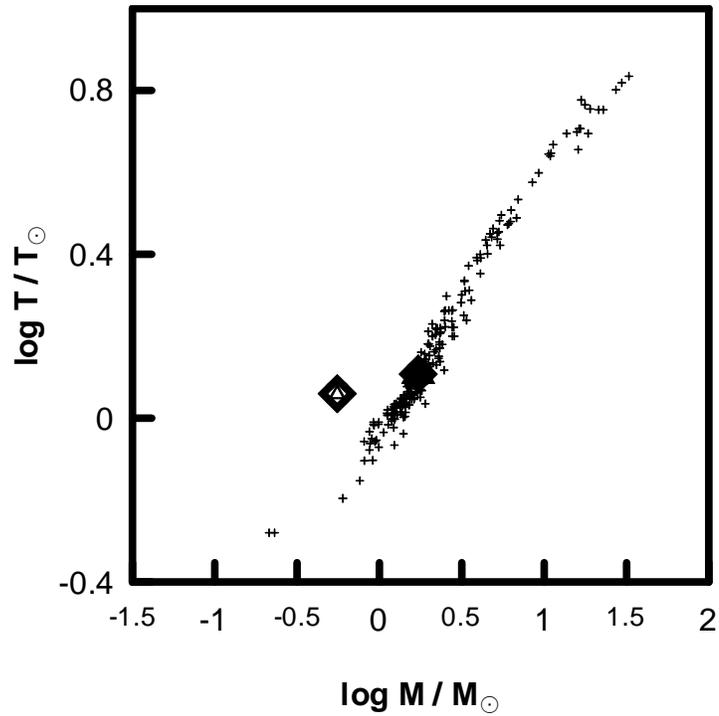
Figure 7: Position of the components of DK Cyg on the empirical mass-Teff relation for low-intermediate mass stars by Malkov (2007).

**Acknowledgements**

This research has made use of the **NASA**'s **ADS** and the available on-line material of the **IBVS**. Our sincere thanks to Dr. Bob Nelson, who allowed us to use his windows interface code WDwint56a and for his helpful discussions and advices.

# References

Awadalla, N., 1994, A&A.,289, 137.

Baldwin, M., Samolyk, G., 2003, AAVSO, No. 8.

Baran, A.; Zola, S.; Rucinski, S.; Kreiner, J.; Drozdz, M., 2004, Acta Astron., 54, 195.

Binnendijk1964, A.J.,69, 157.

Borkovits, T., Elkhateeb, M.; Csizmadia, S.; Bíró, I.; Hegedüs, T.; and Csorvasi, R. 2005, A&A, 441, 1087.

Borkovits, T.; Bíró, I.; Csizmadia, S.; Patkós, L.; Hegedüs, T.; Pál, A.; Kóspál, Á.; Klagyivik, P.2004, IBVS, 5579.

Borkovits, Tamás; Bíró, Imre Barna; Hegedüs, Tibor; Csizmadia, Szilárd; Kovács, Tamás; Kóspál, Ágnes; Pál, András; Könyves, Vera; Moór, Attila, 2002, IBVS, 5313.

Bradstreet D., Steelman, D., 2002, BAAS, 34, 1224.

Diethelm, R., 2010, IBVS, 5920.

Diethelm, R., 2012, IBVS, 6011.

Diethelm, R., 2013, IBVS, 6042.

Dogru, S., Dönmez, A.; Tüysüz, M.; Dogru, D.; Özkardes, B.; Soydugan, E.; and Soydugan, F.2007, IBVS, 5746.

Dogru, S., Erdem, A.; Aliçavus, F.; Akin, T.; and Kanvermez, Ç., 2011, IBVS, 5988.

Drozdz, M. and Ogloza, W. 2005, IBVS, 5623.

Elkhateeb, M. M., Nouh, M. I., Saad, A. S. (2014), Submitted to Research in Astronomy and Astrophysics.

Elkhateeb, M. M. and Nouh, M. I. (2014), Submitted to Astrophysics and Space Science.

Erkan, N., Erdem, A.; Akin, T.; Aliçavus, F.; Soydugan, F.2010, IBVS, 5924.

Gerner, H., 2008, JAAVSO, 36.

Girardi, L., Bressan, A., Bertelli, G., Chiosi, C., 2000, A&AS 141, 371.

Guthnick, P., Prager, R., 1927, Kl.Ver.B.-Babelsberg, 4.

Hinderer, F., 1960, J. des Observateurs, 43, 161.

Hubscher, J., 2006, IBVS, 5731.

Hubscher, J., Agerer, F.; Wunder, E. 1990, B.A.V.Mitt.,59.

Hubscher, J., Steinbach, H., Walter, F., 2008, IBVS, 5830.

Hubscher, J., Lehmann, P.; Monninger, G.; Steinbach, H.; Walter, F., 2010, IBVS, 5941..

Kiss, L, Kaszás, G., Fürész, G., Vinkó, J.1999, I.B.V.S., 4681.

Kreiner, J., Kim, C., Nha, I., 2001, An Atlas of (O-C Diagrams of Eclipsing Binary Stars,

Krakow: Wydawn. Nauk. Akad. Pedagogicznej.

Kukarkin, B., Kholopov, P., Efremove, Y., et al. 1970, GCVS, Nauka, Mosco.

Lucy, L., 1967, Z.F. Astrophys. 65, 89.

Maceroni, C.; van't Veer, F. (1996), A&A, 311, 523.

Malkov, O. Yu, 2007, MNRAS, 382, 1073.

Menzies, K. 2009, JAAVSO, 37

Mochnacki, S., Doughty, N., 1972, MNRAS, 156, 243.

Nelson, R., 2009, http://members.shaw.ca/bob.nelson/software1.htm

Paparo, M., Hamdy, M., Jankovics, I, 1985, IBVS, 2838

Piotrowski, S., 1936, Acta Astr.ser.c, 2, 154

Rucinski, S., 1969, Acta Astron. 19, 156.

Rucinski, S., Lu, W., 1999, A.J, 118, 2451.

Samolyk, G. 2009, JAAVSO, 37

Samus, N., Durlevich, O., Goranskij, V., Kazarovets, E, Kireeva, N., Pastukhova, E., Zharova,
    A., (2014), http://www.sai.msu.su/gcvs/cgi-bin/search.htm

Sarounová, L., Wolf , M., 2005, I.B.V.S., 5594

Simmons, N., 2011, www.aavso.org/data-download.

Szafraniec, R 1953, Acta Astr.ser.c, 5, 51

Tsesevitch, V., 1954, Odessa Izv.,4,258.

Van Hamme, W., 1993, AJ, 106, 2096.

Wolf, M., Molik, P., Hornoch, K, Sarounova, L., 2000, A&ASS,147, 243.